\documentclass[sigconf]{acmart}

\usepackage[boxed]{algorithm2e}
\usepackage{bm}
\usepackage{soul}
\usepackage{subfig}
\usepackage{graphicx} %use graph format
\usepackage{multirow}
\usepackage{booktabs}
\usepackage{balance}

\DeclareMathOperator*{\argmin}{arg\,min}

\AtBeginDocument{%
	\providecommand\BibTeX{{%
			\normalfont B\kern-0.5em{\scshape i\kern-0.25em b}\kern-0.8em\TeX}}}

\copyrightyear{2020}
\acmYear{2020}
\acmConference[WWW '20]{Proceedings of The Web Conference 2020}{April 20--24, 2020}{Taipei, Taiwan}
\acmBooktitle{Proceedings of The Web Conference 2020 (WWW '20), April 20--24, 2020, Taipei, Taiwan}
\acmPrice{}
\acmDOI{10.1145/3366423.3379992}
\acmISBN{978-1-4503-7023-3/20/04}

\begin{document}
	
\title{Practical Data Poisoning Attack against Next-Item Recommendation}
% \iffalse
% 
\author{Hengtong Zhang}
\authornote{This work was done when the first author was an intern at Alibaba Group.}
\affiliation{%
	\institution{State University of New York at Buffalo}
}
\email{hengtong@buffalo.edu}
\author{Yaliang Li}
\affiliation{%
	\institution{Alibaba Group}
}
\email{yaliang.li@alibaba-inc.com}
\author{Bolin Ding}
\affiliation{%
	\institution{Alibaba Group}
}
\email{bolin.ding@alibaba-inc.com}
\author{Jing Gao}
\affiliation{%
	\institution{State University of New York at Buffalo}
}
\email{jing@buffalo.edu}

%\renewcommand{\shortauthors}{Zhang, et al.}
% \fi	
\begin{abstract}
Online recommendation systems make use of a variety of information sources to provide users the items that users are potentially interested in. However, due to the openness of the online platform, recommendation systems are vulnerable to data poisoning attacks.
Existing attack approaches are either based on simple heuristic rules or designed against specific recommendations approaches.
The former often suffers unsatisfactory performance, while the latter requires strong knowledge of the target system. 
In this paper, we focus on a general next-item recommendation setting and propose a practical poisoning attack approach named \textit{LOKI} against blackbox recommendation systems. 
The proposed \textit{LOKI} utilizes the reinforcement learning algorithm to train the attack agent, which can be used to generate user behavior samples for data poisoning. 
In real-world recommendation systems, the cost of retraining recommendation models is high, and the interaction frequency between users and a recommendation system is restricted.
Given these real-world restrictions, we propose to let the agent interact with a recommender simulator instead of the target recommendation system and leverage the transferability of the generated adversarial samples to poison the target system. We also propose to use the influence function to efficiently estimate the influence of injected samples on the recommendation results, without re-training the models within the simulator. 
Extensive experiments on two datasets against four representative recommendation models show that the proposed \textit{LOKI} achieves better attacking performance than existing methods.
\end{abstract}

%\begin{CCSXML}
%<ccs2012>
%  <concept_id>10010520.10010553.10010562</concept_id>
%  <concept_desc>Computer systems organization~Embedded systems</concept_desc>
%  <concept_significance>500</concept_significance>
%  <concept_id>10010520.10010575.10010755</concept_id>
%  <concept_desc>Computer systems organization~Redundancy</concept_desc>
%  <concept_significance>300</concept_significance>
%  <concept_id>10010520.10010553.10010554</concept_id>
%  <concept_desc>Computer systems organization~Robotics</concept_desc>
%  <concept_significance>100</concept_significance>
%  <concept_id>10003033.10003083.10003095</concept_id>
%  <concept_desc>Networks~Network reliability</concept_desc>
%  <concept_significance>100</concept_significance>
%</ccs2012>
%\end{CCSXML}
%
%\ccsdesc[500]{Computer systems organization~Embedded systems}


\maketitle

\section{Introduction}
In the era of big data, one of the fundamental challenges for web users is information overload, because of which users struggle in locating the information they indeed need. Recommendation systems, which suggest items (e.g., movies, products, music,  etc.) that are likely to interest users based on their historical behaviors, are proposed to alleviate the information overload issue. 
Nowadays, recommendation systems are widely deployed by Web service platforms (e.g., YouTube, Amazon, and Taobao) and play an important role in guiding users to make decisions and choices.

It is commonly assumed that online recommendation systems are honorable and unbiased. They  recommend users the items that match their personal interests.
However, the openness of recommendation systems and the potential benefit of manipulating recommendation systems offer both opportunities and incentives for malicious parties to launch attacks.
Recent studies~\cite{li2016data,o2004collaborative,yang2017fake,mobasher2007toward,lam2004shilling} have demonstrated that recommendation systems are vulnerable to poisoning attacks. In these poisoning attacks,  well-crafted data is injected into the training set of a recommendation system by a group of malicious users. Such poisoning attacks make the system deliver recommendations as attackers desire.

Existing poisoning attacks can be categorized into two types. 
The first type of work is generally based on manually designed heuristic rules. 
For example, \cite{o2004collaborative} design rules that leverage the following intuition: items that are usually selected together by users are treated as highly correlated by recommendation systems. 
To promote a target item to target users, attackers utilize controlled users to fake the co-occurrence between the target item and popular items.
Nevertheless, such heuristic rules are not able to cover various patterns of behavior in the recommendation data.
Therefore, the performances of these attack methods are usually unsatisfactory.
The other line of methods are designed for certain types of recommendation methods like matrix factorization based  models~\cite{li2016data}.
However, the architecture and the parameters of the recommendation systems in real-world platforms are generally unknown to the attackers.
Usually, the only information that the attackers can rely on to infer the characteristics of the recommendation systems is the recommendation results of the users they controlled, and the frequency of these interactions is often limited.
Thus, there is still a noticeable gap before these attacks methods can be deployed in real practice.
% utilized 

In this work, we propose a novel practical adversarial attack framework against sophisticated blackbox recommendation systems. We focus on one of the most common next-item recommendation setting, which aims to recommend top-$K$ potentially preferred items for each user. 
The proposed reinforcement learning based framework \textit{LOKI} learns an attack agent to generate adversarial user behavior sequences for data poisoning attacks. 
Unlike existing attack methods designed for certain types of recommendation methods,  reinforcement learning algorithms can utilize the feedback from the recommendation systems, instead of comprehensive knowledge of architecture and parameters, to learn the agent's policy.
Nevertheless, in practice, the attacker cannot control the target recommendation system to be retrained to get the feedback and update the attack strategy.
In addition, recommendation system service providers generally restrict feedback frequency, but a reinforcement learning based framework requires a large number of feedback to train a policy function. 
Due to this discrepancy, we cannot directly rely on the feedback from the target recommendation system to train a policy within a tolerated time period.

To tackle this challenge, we propose to construct a local recommender simulator to imitate the target model, and let the reinforcement framework get reward feedback from the recommender simulator instead of the target recommendation system.
The local recommender simulator is constructed by constructing an ensemble of multiple representative recommendation models. 
The intuition behind such a design is that if two recommenders can both get similar recommendation results on a given dataset, then the adversarial samples generated for one of the recommenders can be used to attack the other.
Such transferability makes the recommender simulator a good substitute for the target recommendation system in terms of guiding the attack agent. 
Moreover, even with the help of a local simulator, it is still time-consuming to retrain the recommendation systems within the simulator using the contaminated data for attack outcome. 
To alleviate this problem, we design a component named outcome estimator, which is based on the influence function. The outcome estimator can efficiently estimate the influence of the injected adversarial samples on the attack outcomes. These designs ensure that the proposed adversarial attack framework for recommendation systems is practical and effective. 

In the experiments, we adopt representative recommendation models as targets and conduct attacks on a real-world dataset to evaluate the proposed poisoning attack framework. Experimental results show that the proposed \textit{LOKI} outperforms baseline attack methods. We also provide further analysis of the factors that influence the attack outcome.

\section{Threat Model}\label{sec:pre}

To facilitate the discussions in the rest of this paper,  we specify and formulate the next-item recommendation task as follows:

\begin{definition}[Next Item Recommendation]
Let $\mathcal{U}$ be the set of users and $\mathcal{V}$ be the set of items, we use $\bm{x}_u = [x_{u}^1, x_{u}^2, \cdots, x_{u}^{m_u}]$ to denote a sequence of items that user $u$ has chosen before in a chronological order in which $x_{u}^v \subseteq \mathcal{V}$. $m_u$ denotes the number of items chosen by user $u$. Given existing sequences, the goal of the next-item recommendation is to output a $K$-sized ordered item list, which predicts the next item that the user will choose.
\end{definition}

With the aforementioned definition, let us detail the threat model of the attack against the next-item recommendation.

\textbf{Attack Goal}: An attacker's goal is to promote a set of target items 
% $V_{target} = \{v_i\}$ 
to as many target users as possible. Specifically, suppose the system recommends $K$ items to each user,
\emph{the attacker's goal is to maximize averaged display rate, which denotes the fraction of target users whose top-K recommendations results include the target items.}
Note that an attacker could also demote a target item.
% , i.e., minimize the hit rate of the target item. 
Demotion can be viewed as a special case of promotion as an attacker can promote other items such that the target item is demoted in recommendation lists. Thus, in this paper, we focus on promotion attacks.

\textbf{Attack Approach}: 
To achieve the attack goal, we consider the most general scenario in which the attackers can inject controlled users into the recommendation system. These controlled users visit or rate to well-selected items, which are named as \emph{proxy items}, step-by-step. Thus, the well-crafted activities of each controlled user form a \emph{behavior sequence}. 
To make the injection unnoticeable, the number of visits or ratings each controlled user conducts is limited to at most $M$. 

\textbf{The Knowledge and Capability of the Attacker}: In this paper, we assume that the attacker is granted the following knowledge and capability. 
\begin{enumerate}
	\item The attacker can access the full activity history of all the users in the recommendation system. 
	\item The attacker has limited resources so the attacker can merely inject a limited number of controlled users
	which 
	can easily be bought from the underground market\footnote{ https://www.buzzfeednews.com/article/leticiamiranda/amazon-marketplace-sellers-black-hat-scams-search-rankings }.
	\item The attacker does not know the details about the target recommendation system, for instance, the parameters and the architecture of the recommendation model. Such setting is also known as {\emph{blackbox setting}}. 
	\item The attacker can only receive a limited number of feedback (e.g., display rates) from the blackbox recommendation model.
	\item The attacker does NOT know when the target blackbox recommendation model is retrained.
\end{enumerate}

\section{Methodology}
In this section, we first provide an overview of the proposed reinforcement learning based framework. Then we describe the detailed design of each component of the framework.

\begin{figure}[t!]
	\centering
	\includegraphics[width=\linewidth]{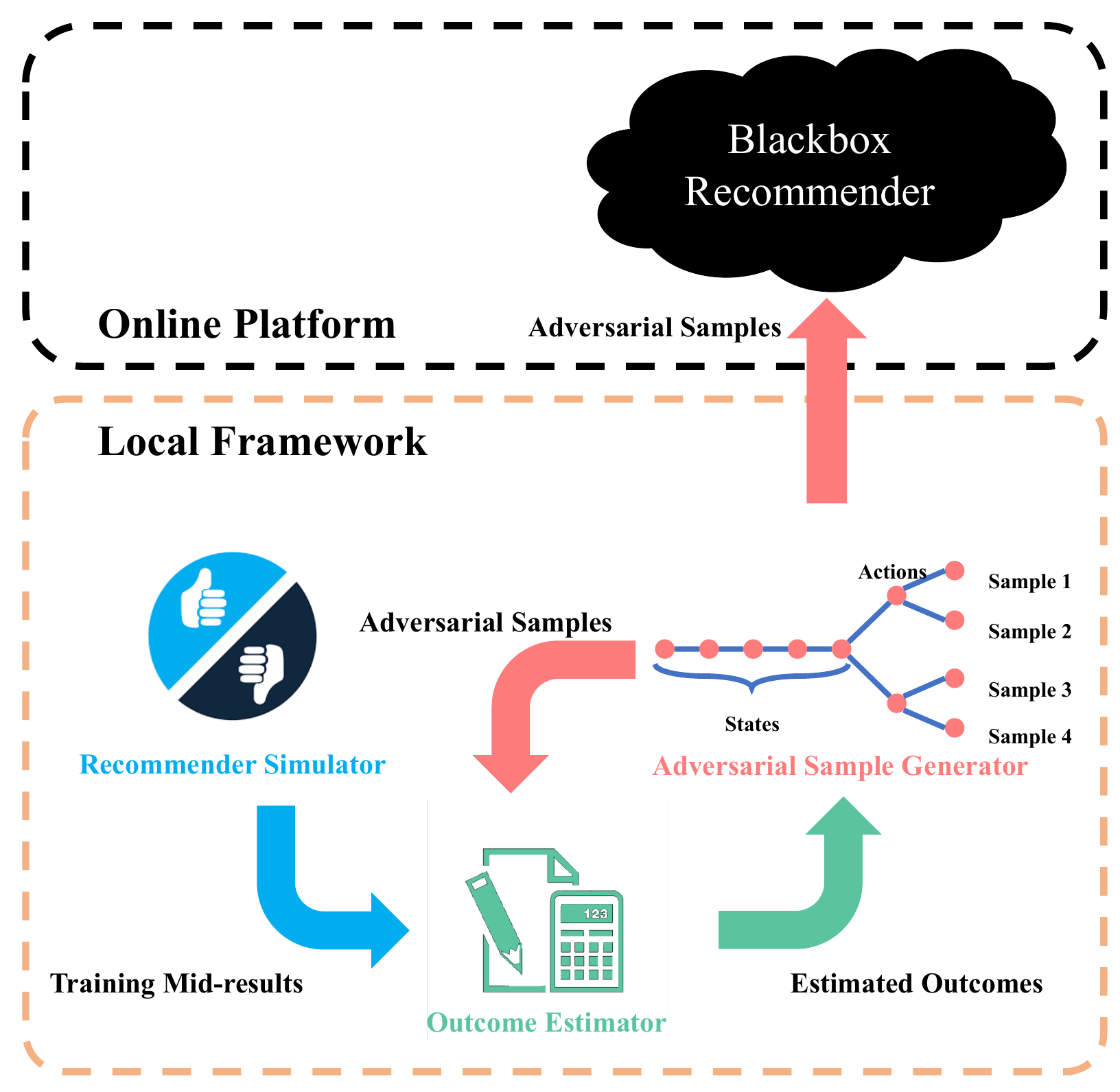}
	\caption{Overview of the proposed framework LOKI.}
	\label{fig:framework}
\end{figure}

\subsection{Framework Overview}\label{sec:frame}

Intuitively, data poisoning can be regarded as the creation of new sequential patterns that involve the target items in the training set of the target recommendation system.
In a crafted sequential adversarial sample, the user behavior history is inherently crucial in determining the next behavior. These sequential adversarial samples together contribute to the manipulation goal.
Generating adversarial samples is essentially a multi-step decision process, in which the generator ought to select specific actions for the controlled users to maximize attack outcome. This fits the reinforcement learning setting.
From the perspective of reinforcement learning, the goal is to learn a policy function to generate sequential adversarial user behavior samples, which can maximize the averaged display rate of the target users.

Based on the aforementioned motivation, we propose a reinforcement learning based framework to learn the policy function.  The overall architecture of the proposed  framework LOKI is illustrated in Figure~\ref{fig:framework}. The target blackbox recommendation system is deployed on an e-commerce platform.
\textcolor{black}{The proposed framework consists of three components: (1) recommender simulator, (2)  outcome estimator, and (3) adversarial sample generator. In the following sections, we describe the details of these components one-by-one.}

\subsection{Recommender Simulator}
The idea of constructing surrogate models and utilizing the transferability property of adversarial samples to attack the target machine learning models is adopted by multiple attack approaches~\cite{biggio2013evasion,papernot2017practical,demontis2019adversarial}.
In this paper, the proposed recommender simulator simulates the recommendation preference of the target model. 
The simulator consists of multiple separated recommendation models, which are trained on the same dataset.  
Recommendation results from these models are aggregated via weighted voting.
Suppose $M$ different recommendation models are deployed to recommend items for user $u$ and the rank of item $i$ in the $m$-th model is denoted as $rank_m(i)$. \textit{The higher item $i$ ranks, the smaller $rank_m(i)$ is.} Here, we define the preference score of item $i$ in the simulator via Eq.~\eqref{eq:score}. All the items are then ranked according to this scoring function:
\begin{equation}
    score(i) = -\frac{1}{M}\sum_{m=1}^M w_m \cdot rank_m(i),
    \label{eq:score}
\end{equation}
where $w_m$ stands for the weight of the $m$-th recommendation model. Ideally, these weights are used to adjust the simulator to match the characteristics of the target recommender. 

\subsection{ Outcome Estimator}\label{sec:est}

As mentioned in Section~\ref{sec:frame}, we need to utilize the manipulation outcome of the current adversarial samples as reward feedback to update the policy network of the adversarial sample generator.
The most straightforward way to obtain the outcome is to retrain the entire model. However, retraining the online recommendation system is prohibitively slow (from a few hours to days for a single retraining). To make the attack methodology practical, we 
propose to use influence function for an efficient estimation of the manipulation outcome, motivated by robust statistics. 

Formally speaking, the parameter estimator of the recommendation models on the clean dataset is: $
\hat{\theta} := \argmin_{\theta} \frac{1}{N}\sum_{i=1}^N \mathcal{L} (z_i; \theta),    $
where $\theta$ denotes the parameter vector, $\mathcal{L}$ stands for the loss function of the recommendation model. $z_i$ denotes a sample in the dataset, and $N$ stands for the total number of samples in the training set. For collaborative filtering models, a sample is a single user-item pair $(u, v)$. For session-based recommendation models, given the behavior sequence $\bm{x}_u = [x_{u}^1, x_{u}^2, \cdots, x_{u}^m]$ of a user $u$, each training sample is made up of a subsequence and the ground truth next item, i.e., $([x_{u}^1], x_{u}^2)$, $([x_{u}^1, x_{u}^2], x_{u}^3), \cdots$, $([x_{u}^1, x_{u}^2, \cdots, x_{u}^{n-1}],x_{u}^n)$.

Now let us move on to the discussion of the influence function.
Suppose we upweight a sample $z_\delta$ by a small $\epsilon$ in the training set, the new estimation of $\theta$ is given as: 
$ \hat{\theta}_{z_\delta} := \argmin_{\theta} \frac{1}{N}\sum_{i=1}^N \mathcal{L} (z_i; \theta)   + \epsilon \mathcal{L} (z_\delta; \theta).$
When $\epsilon \rightarrow 0$, according to the classic results in~\cite{cook1982residuals}, the influence of upweighting $z_\delta$ on the parameter $\theta$ is given by:
\begin{equation}
\frac{d \hat{\theta}_{z_\delta}}{d \epsilon} = \hat{\theta}_{z_\delta} - \hat{\theta} \approx -H^{-1}_{\hat{\theta}} \nabla_\theta \mathcal{L}(z_\delta; \hat{\theta}),
\end{equation}
where 
$
H_{\hat{\theta}} := \frac{1}{N} \sum_{i=1}^{N} \nabla_\theta^2 \mathcal{L}(z_i, \theta),    
$
denotes the Hessian matrix of the loss function. Given the fact that the number of users is large in the recommendation datasets, injecting a data sample is the same as upweighting the sample by $\epsilon \approx \frac{1}{N}$. 

Here, the key computation bottleneck lies in the calculation of the huge  inverse Hessian matrix $H_\theta^{-1}$. Given a sample $z_j$, we use implicit Hessian-vector products (HVPs) \cite{koh2017understanding,agarwal2017second} to efficiently approximate $-H_\theta^{-1} \nabla_\theta L(z_j, \hat{\theta})$. 

Based on an approximate estimate of the sample upweight's influence on parameter $\hat{\theta}$, we further calculate the influence on the prediction scoring function w.r.t. the perturbation. Specifically, suppose we want to promote an product $v'$ to user $u'$, we can treat this as a \textit{target sample} $z^{test}_{u'v'}$in the test set. The influence on the prediction scoring function w.r.t. can be written as:
\begin{equation}
\begin{split}
    \frac{d f_{test}(z^{test}_{u'v'}; \hat{\theta})}{d \epsilon} =& \frac{d f_{test}(z^{test}_{u'v'}; \hat{\theta}) }{d \hat{\theta}_{z_\delta}} \cdot \frac{d \hat{\theta}_{z_\delta}}{d \epsilon}\\
    \approx& -\nabla_\theta f_{test}(z^{test}_{u'v'}; \hat{\theta}) H^{-1}_{\hat{\theta}} \nabla_\theta \mathcal{L}(z_\delta; \hat{\theta}),
\end{split}
\label{eq:diff}
\end{equation}
where $f_{test}$ is the prediction scoring function used by the recommender system in the test phase. This result is further used to design rewards for efficient agent policy training.

\subsection{Adversarial Sample Generator}\label{sec:rl}

The adversarial attack against a local recommender simulator is essentially interpreted as a multi-step decision problem.
In this section, we translate this decision problem into a Markov Decision Process (MDP).
MDP is defined as a tuple $(\mathcal{S}, \mathcal{A}, \mathcal{P}, \mathcal{R}, \gamma)$, where $\mathcal{S}$ is a set of states, $\mathcal{A}$ is a set of actions,
$\mathcal{P}$ is the transition probabilities, $\mathcal{R}$ is the immediate reward, and $\gamma$ is the discount factor. 
In the context of this paper, the MDP can be specified as follows:

\begin{itemize}
\item 
\textbf{Action space $\mathcal{A}$}: As mentioned in Section~\ref{sec:frame}, the attacker determines specific items organized in a proper sequence for each controlled user. Instead of taking the set of all the  possible items as action space, we divide the item set into groups and use the set of all the groups as action space. The main reason for using coarse groups instead of items for action space is due to the concern in learning efficiency. 
Learning action strategies for every single item is not only costly but also unnecessary for the attack goal. This is because adversarial samples do not need to follow the exact sample pattern.
Here we define one of the groups as the collection of all the target items, one of the groups as the collection of all the items already selected by the target users. The remaining groups are obtained by item clustering, in which each group represents items with similar properties.
This item clustering takes the feature vectors of all the items and an integer $c$ as input and divides the items into $c$ clusters. Here, we utilize non-negative matrix factorization~\cite{lee2001algorithms} to extract item features and use K-means~\cite{macqueen1967some} algorithm for clustering.
\begin{figure}[t]
    \centering
    \includegraphics[width=\linewidth]{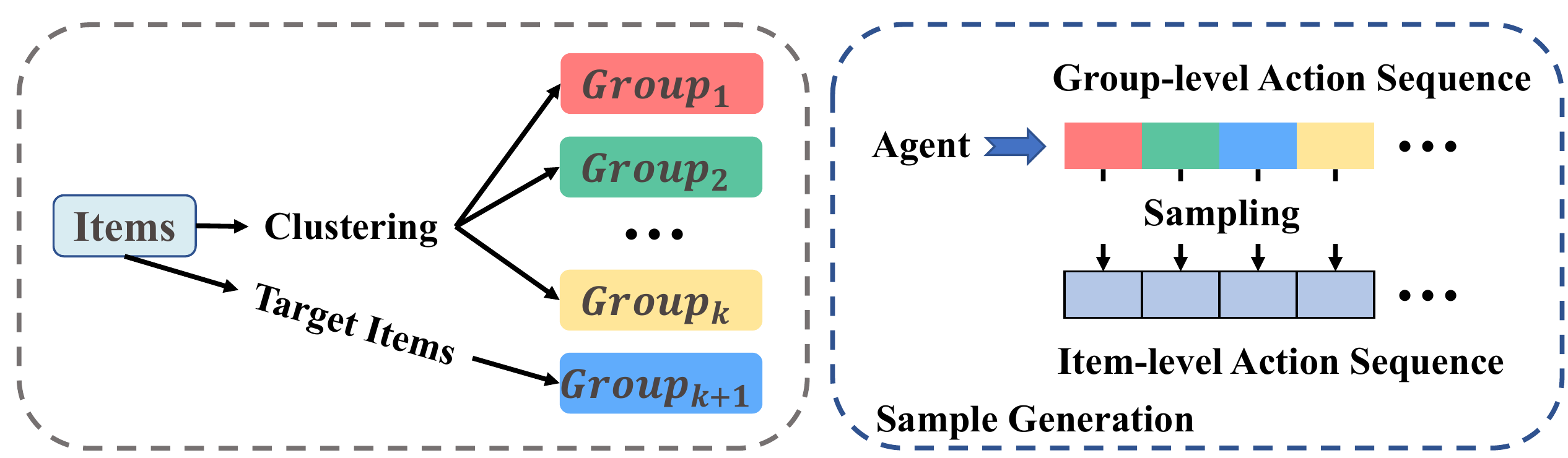}
    \caption{Generation of adversarial samples.}     \vspace{-0.1in}
   \label{fig:action}
\end{figure}
After item groups are obtained, during the phases of training and testing the agent, \textit{group-level actions} are sampled step-by-step from the policy, forming a \textit{group-level action sequence}.
Then sequential poisoning samples are sampled step-by-step from the corresponding group indicated by the current step of the \textit{group-level action sequence}. This process is illustrated in Figure~\ref{fig:action}. The left side of Figure~\ref{fig:action} shows the process of clustering items into groups and the right side illustrates the process of generating the poisoning samples.
\item 
\noindent\textbf{State $\mathcal{S}$}: The state is defined as the {action subsequence} before current step $t$ and the actions all come from the action space mentioned above.
\item
\noindent\textbf{Reward $\mathcal{R}$}: As aforementioned, the purpose of the attacker is to manipulate the local recommender simulator and further the target recommender. Hence, the RL framework should learn a policy that promotes the estimated prediction scores of target items given by the target users as much as possible. Thus, we design the reward as \textit{the weighted averaged influence on the prediction scoring function, i.e., Eq.~\eqref{eq:diff}, of all the target samples}. The weights are assigned manually to indicate the importance of each recommendation simulator.
\end{itemize}

Here, we apply Deep Q-Network (DQN) to estimate the action-value function. 
The representation of the existing sequence, i.e., state, is modeled via a GRU (Gated Recurrent Unit) layer, and the representation of each type of \textit{actions} is extracted via a embedding layer. Finally, we deploy a fully connect layer which takes the final output of the GRU layer as input and output the estimated action-values.

The DQN is trained via an iterative algorithm. In each iteration, there are two stages, replay memory generation stage and parameters update stage. In replay memory generation stage, the agent generates a group-level action $a_t$ according to an $\epsilon$-greedy policy and current state $s_t$. Then the item-level sequences are generated by sampling items from the corresponding group suggested by each step in the group-level sequence.
After that, the agent observes the reward $r_t$ from the outcome estimator and updates the state. 
For parameter update stage: the agent samples a $(s_t, a_t, r_t, s_{t+1})$ from replay memory, and then updates the parameters.

\section{Experiments}
In this section, we test the proposed \textit{LOKI} on real against different recommendation methods. The experimental results show that the proposed method outperforms existing attack strategies.  Besides, we systematically study the effect of some key factors. 

\subsection{Datasets} 
To demonstrate the performance of the proposed poisoning attack framework, we adopt the \emph{Amazon Beauty}, which is one category of 
the widely used recommendation dataset series named \textit{Amazon}~\cite{he2016ups}.
The dataset used in this paper mainly focuses on hair and skin care products and is extracted from large corpora of product reviews crawled from \textit{Amazon.com}.
The number of users and items in the dataset are 22,363 and 12,101, respectively. The number of total user activities (i.e., purchase and review) is 146,031. On average, each user is involved in 6.53 activities and each item is involved in 12.06 activities. We followed the similar preprocessing procedure introduced in~\cite{tang2018personalized,kang2018self} and filter out the users with less than five activities and items with less than five feedbacks.

\subsection{Experimental Settings}

\subsubsection{Baseline Attack Methods} 

As aforementioned, there is no existing work solving exactly the same task considered in this paper. 
Although there are some existing attack approaches~\cite{li2016data} against recommendation methods, they are mostly designed for \textit{whitebox setting} and require a strong knowledge of the architecture and parameters of the corresponding model. Therefore, these methods cannot be used in the \textit{blackbox setting} discussed in this paper. Hence, we compare the proposed \textit{LOKI} with several existing heuristic-based attack strategies.
\begin{itemize}
    \item \textbf{None}: This denotes the circumstance when no attack is conducted.
	\item \textbf{Random}: In this baseline method, the attacker mixes the target items and the randomly picked items to form a repository for each controlled account. In each step, the controlled user picks items at random from the item repository without repetition.
	\item \textbf{Popular}: This is a variant of ~\cite{yang2017fake}. In this baseline method, attackers inject fake co-visitations between the popular items and the target items, to promote the target items.
\end{itemize}

\subsubsection{Target Recommendation Methods}

In this section, we consider the following target methods for performance comparisons. The parameters of these target methods are set following the suggestion in the original papers. 
\begin{itemize}
	\item \textbf{BPRMF}~\cite{rendle2009bpr} is a factorization based personalized ranking approach. It is a state-of-the-art method for non-sequential item recommendation on implicit feedback data.
	\item \textbf{FPMC}~\cite{rendle2010factorizing} is a classic hybrid model combing Markov chain and matrix factorization for next-basket recommendation. FPMC can model the user's long-term preference and the short-term item-to-item transition.
\end{itemize}
 For each user $u$ in the dataset, suppose the length of $u$'s sequence is $T_u$, we hold the first $T_u-2$  actions in the sequence as the training set and use the next one action as the validation set to search for the optimal hyperparameter settings for these recommendation models. The attack methods aim to manipulate the prediction of the next item, i.e. item $T_u$.\\

To simulate the interactions between the target blackbox recommendation system and the recommender simulator, we adopt the ``leave-one-out strategy''. 
That is to say we use a specific recommendation model as the target, which is blind to the attacker, and use the aggregation of all the methods as the recommender simulator. 
The number of target items and target users in this paper are both fixed to be 20.

\subsubsection{Evaluation Metric} We use the averaged \textit{display rate}, which denotes the fraction of target users whose top-K recommendation results include the target items, as our evaluation metric. The larger the display rate is, the better the attack approach performs.

\begin{figure}[h]
	\centering
	\includegraphics[width=0.25\textwidth]{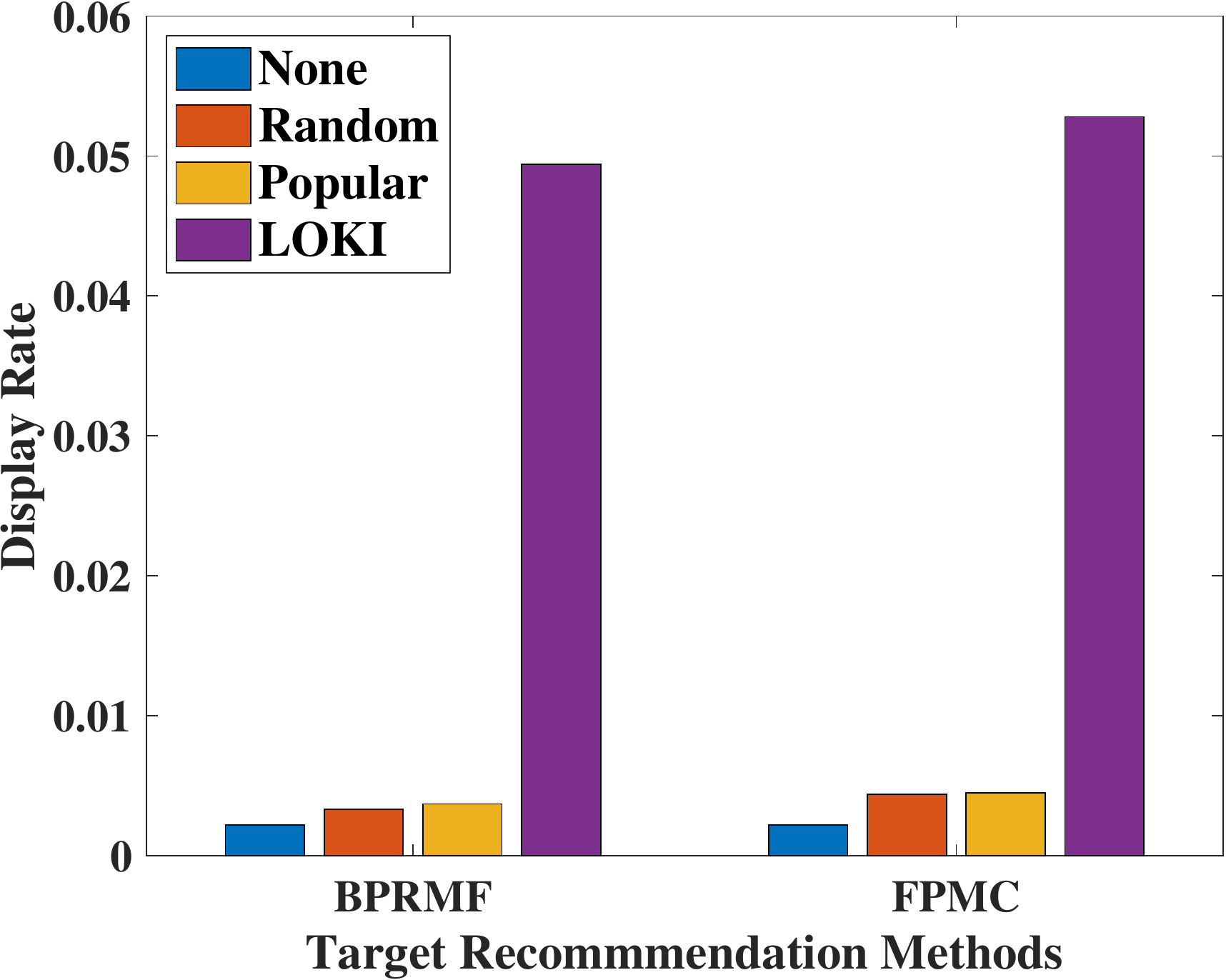}
    \caption{Overall performance of all the attack methods. (Best viewed with color). }
    \label{fig:main}
\end{figure}

\subsection{Result Analysis}

Figure~\ref{fig:main} summarizes the results for all the attack methods. 
Here, we fix the percentage of controlled users to $3$\% and the number of actions per user to $15$. The number of returned items is fixed to 10.
In terms of attack outcome, the proposed \textit{LOKI} achieves the best performance and the improvement is significant.
For example, on compared with the best baseline, the proposed \textit{LOKI}'s display rate increases by over eight times on average.
Among the baseline methods, \textit{Random} simply lets the target items occur in the poisoning sequences without actually creating any new pattern that favors the recommendation of the target items. Thus, \textit{Random} has the worst performance.
\textit{Popular} fakes the co-visitations between the popular items and the target items without considering whether these popular items indeed overlap with the preferences of the target users. Thus, they cannot get a satisfactory performance too.

Compared with these baselines, the proposed \textit{LOKI} takes advantage of the feedback from the local simulator to train an attack agent.
The learned agent is capable of creating more complex patterns to achieve data poisoning goals.
We also notice that the more complicated the target model is, the higher increase in performance metric the proposed \textit{LOKI} achieves. 
For instance, the performance gap is larger when attacking FPMC than attacking BPRMF.
This is because advanced methods are able to capture more complicated user patterns within user behavior sequences. The capability to capture various user patterns leads to better prediction performance in general, but at the same time, 
enables more room for the attack improvement by the proposed \textit{LOKI} with its ability of creating new patterns to poison the recommendation.
That is to say, in some circumstances, the proposed \textit{LOKI} can create certain sequential patterns. However, since relatively simple methods cannot capture these crafted patterns, these methods are less sensitive to the adversarial samples generated by the proposed \textit{LOKI}. 

\begin{figure}
	\centering
	\subfloat [{\small VS BPRMF}] {
		\includegraphics[width=0.22\textwidth]{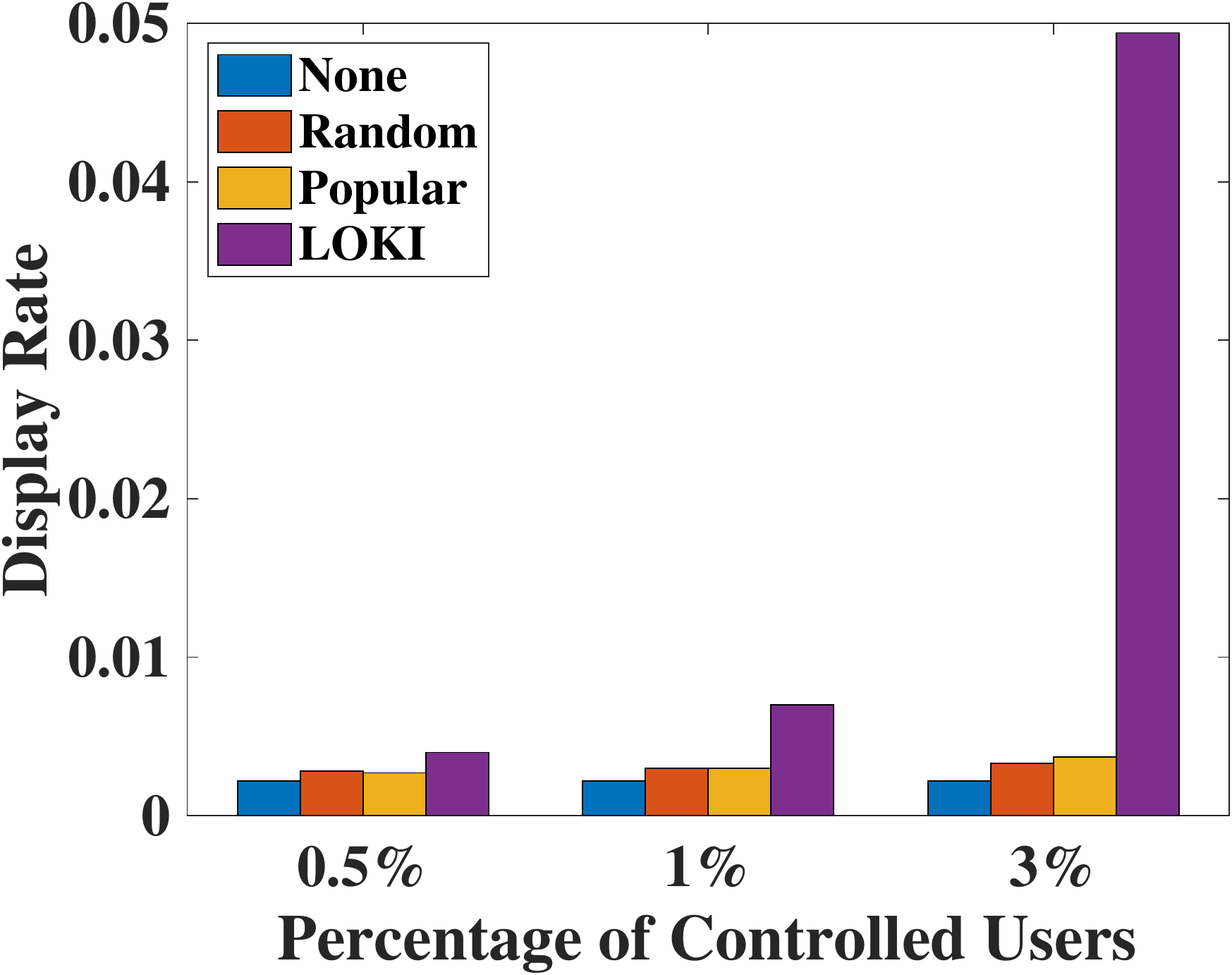}		
	}
	\hfil
	\subfloat [{\small VS FPMC}]{
		\includegraphics[width=0.22\textwidth]{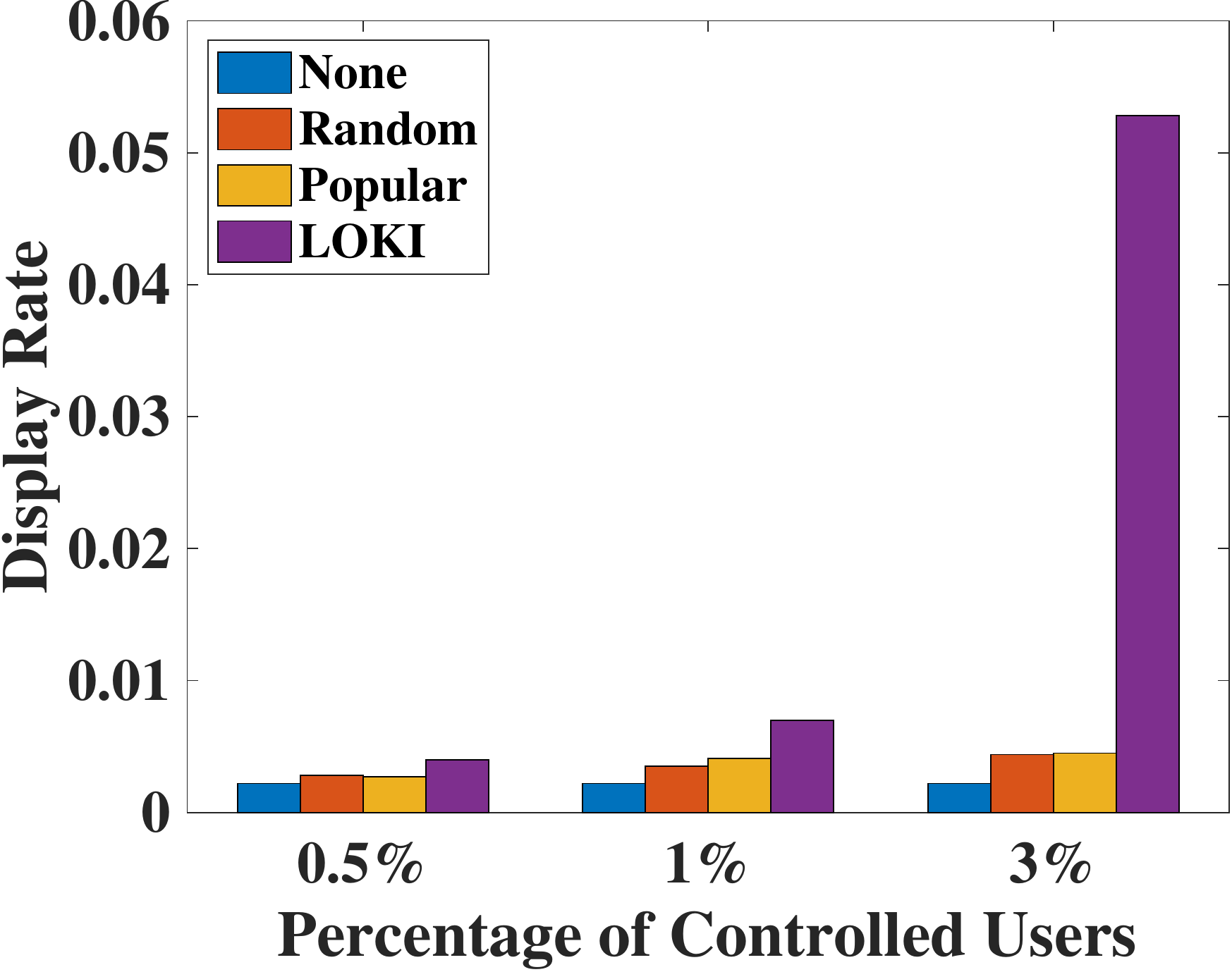}
	}
% \vspace{-0.1in}
    \caption{Impact of the percentage of the controlled users. (Best viewed with color).}
    \label{fig:ratio_ctrl}
\end{figure}

\begin{figure}
	\centering
	\subfloat [{\small VS BPRMF}] {
		\includegraphics[width=0.22\textwidth]{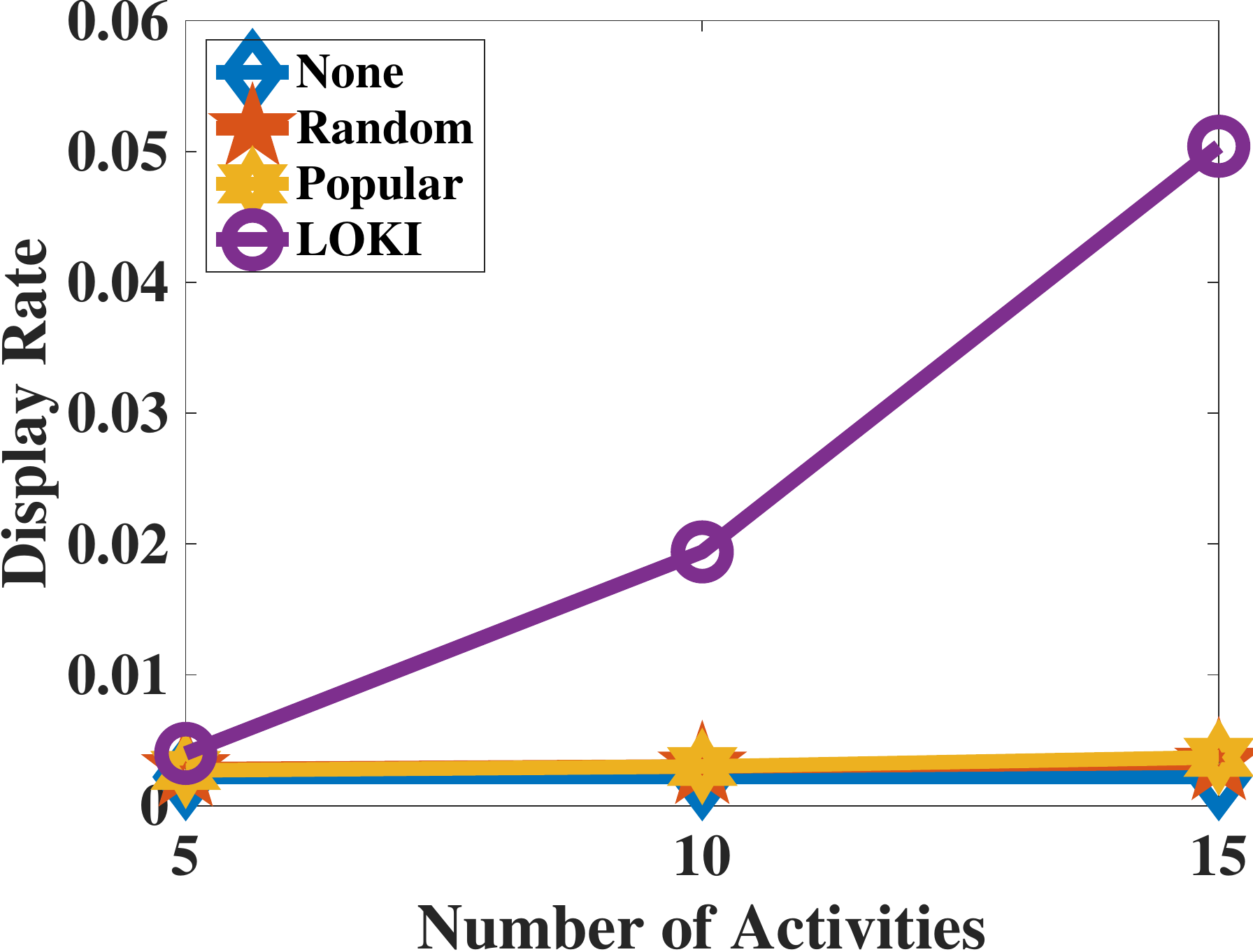}
	}
	\hfil
	\subfloat [{\small VS FPMC}]{
		\includegraphics[width=0.22\textwidth]{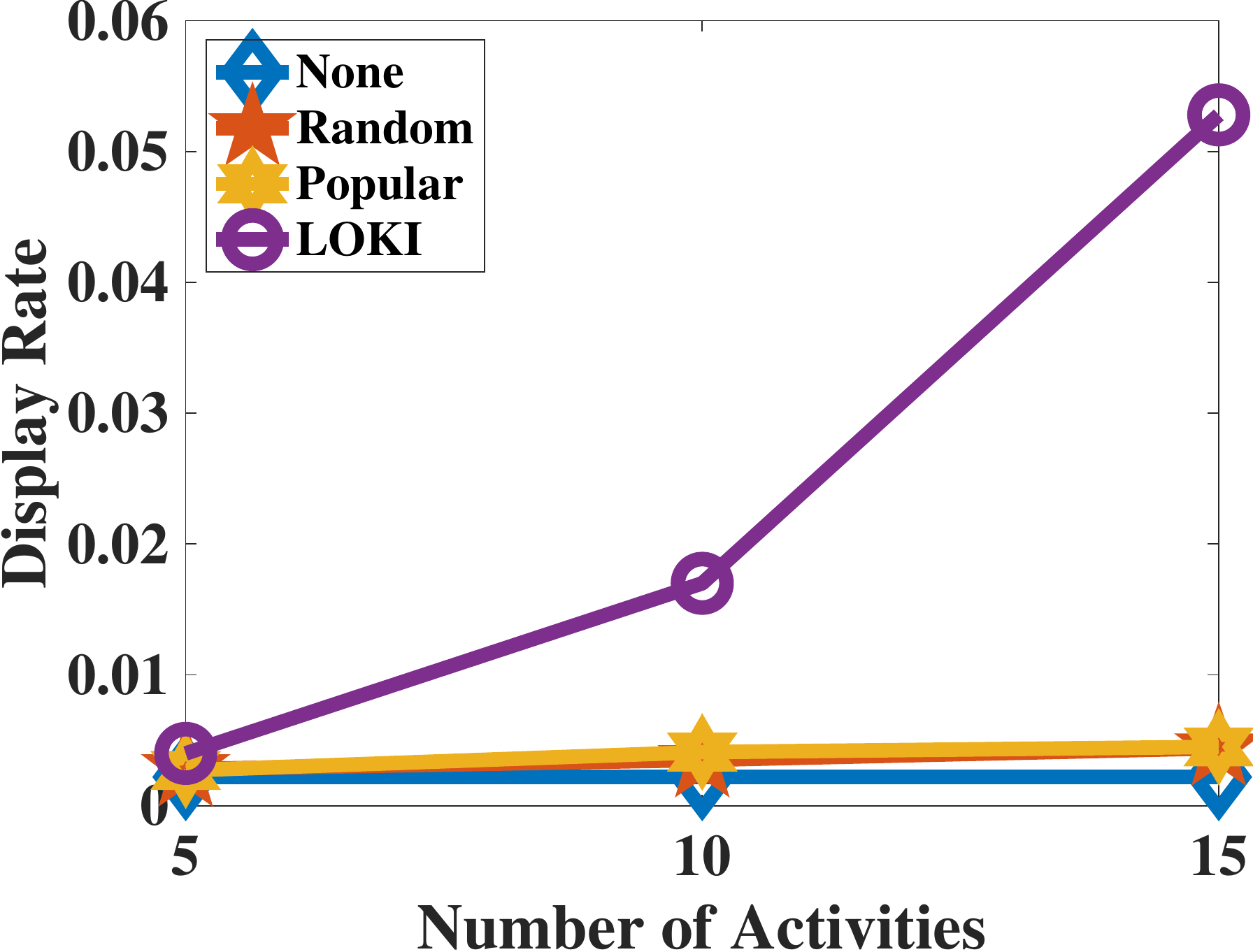}
	}
% \vspace{-0.1in}
    \caption{Impact of the number of activities per controlled user. (Best viewed with color).}
    \label{fig:ratio_steps}
\end{figure}

\subsection{Parameter Analysis}
After discussing the overall experimental results and the characteristics of vulnerable users, we demonstrate the impact of two attack budgets, i.e. (1) the percentage of controlled users recruited by the attacker; (2) the number of activities that each controlled user conducts.

\textbf{Impact of the Percentage of the Controlled Users}. In this experiment, we consider the case where the percentage of the controlled user is low, and evaluate the performance of \textit{LOKI} when this percentage is varied. Here ``\textit{percentage}'' is calculated as the number of the controlled users over the number of normal users. The number of activities per controlled user is fixed to {15} and the recommendation system returns top {10} items. The \textit{display rate} for the real-world datasets is shown in Figure~\ref{fig:ratio_ctrl}. From the figure, we can clearly see that the proposed \textit{LOKI} outperforms the baselines in all cases and can successfully promote the target items.
For instance, when attacking against FPMC, the display rate increases to \textcolor{black}{0.055} even when the percentage of the controlled users is as low as 3\%. Thus, we can conclude that the attack proposed in this paper is effective even with a scant attack budget.

\begin{figure}
	\centering
	\includegraphics[width=0.28\textwidth]{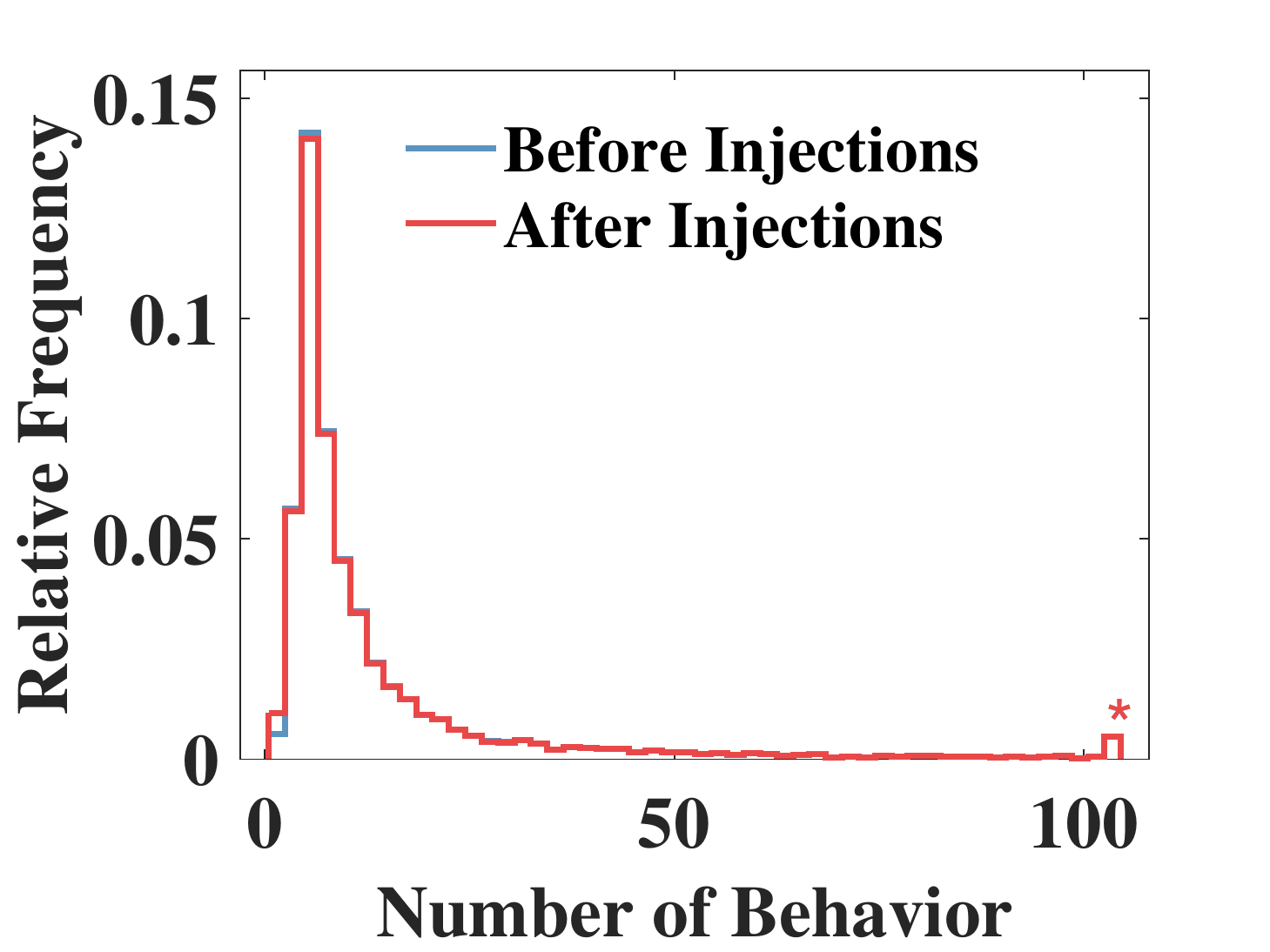}
%	\vspace{-0.1in}
    \caption{Distributions of the length of sequential user behavior samples before and after the injection on Amazon Beauty dataset. (Best viewed with color).}
    \label{fig:dist}
\end{figure}

\noindent\textbf{Impact of the Number of Activities per Controlled User}.
When the percentage of controlled users is given, the number of activities each controlled user conducts is another important attack factor. In this experiment, we fix the percentage of the controlled users to be 3\% and vary the number of activities each controlled user conducts from 5 to 15 for all the datasets. The recommendation system returns top 10 items. The results are shown in Figure~\ref{fig:ratio_steps}. These results show that the proposed \textit{LOKI} outperforms the baseline methods in all cases. With the number of activities per user increasing, the display rates also increase. 
This is because a larger number of activities grant the controlled users more capability to inject the manipulated bias information to the system.
If we look at the distribution of the length sequential user behavior samples in the Amazon dataset, for example, the distribution derived from the dataset in Figure~\ref{fig:dist}, we can see that the distributions before and after the injection are similar.
Hence, injecting such generated sequential samples into the dataset is unnoticeable from the perspective of the online platform operators in practice. 

\section{Related Work}
In this section, we survey the related work from two aspects: general data poisoning attacks and poisoning attacks against recommendation systems.

\textbf{Data Poisoning Attacks}: Data poisoning attacks against machine learning algorithms
have become an important research topic in the field of adversarial machine learning.
This type of attack takes place during the training stage of machine learning models. The attacker tries to contaminate the training data by injecting well-designed samples to force a nefarious model on the learner. Data poisoning attacks have been studied against a wide range of learning systems
%. The first work in this field~\cite{biggio2012poisoning} investigated the vulnerability of support vector machines (SVM), where an attacker progressively injects malicious data points to the training set in order to maximize the classification error.
% Later, Mei et al.~\cite{mei2015using} generalized these attacks against SVM into a bilevel optimization framework against general offline learners with a convex objective function.
% Recently, poisoning attacks have been analyzed on many important machine learning algorithms, 
including
\textcolor{black}{SVM~\cite{biggio2012poisoning}}
neural networks~\cite{munoz2017towards,koh2017understanding}, latent Dirichlet allocation~\cite{mei2015security}, matrix factorization-based collaborative filtering~\cite{li2016data} and autoregressive models~\cite{alfeld2016data,chen2019optimal}.
Existing work has almost exclusively focused on (1) whitebox settings, where the attacker observes the model architecture; (2) continuous data like image or acoustic data. In this paper, we focus on a more challenging blackbox setting, where the attacker does not know the architecture or the parameter of the target model. Instead, the attacker is merely given scant implicit feedback from the target model. 

\textbf{Poisoning Recommendation Systems}: Similar to general data poisoning attacks, poisoning recommendation systems aims to spoof a recommendation system via injecting adversarial samples, such that the system recommends as the attacker desires. The first study on poisoning recommendation systems~\cite{o2004collaborative} was carried out more than a decade ago. In early work, the proposed attacks are usually heuristics-driven.
% and are not optimized to a particular type of recommendation systems. 
For instance, in random attacks~\cite{lam2004shilling}, the attacker randomly selects some items for each injected controlled user and then generates a rating score for each selected item from a normal distribution, whose mean and variance are the same as those of the uncontaminated dataset. These methods rely on user-item ratings which do not exist in the next-item recommendation setting.
Poisoning attacks~\cite{li2016data,yang2017fake} that were proposed recently generate fake behavior that is optimized according to a particular type of recommendation system. Specifically, Li et al.~\cite{li2016data} proposed poisoning attacks for matrix-factorization-based recommendation systems. Yang et al.~\cite{yang2017fake} proposed poisoning attacks for association-rule-based recommendation systems, in which each user injects fake co-visitations between items instead of fake rating scores of items. Unlike these methods, the framework proposed in this paper does not require the details of the target system as prior knowledge. Hence, the proposed framework can be used in a broader spectrum of contexts.

\section{Conclusions}
In this work, we propose a data poisoning attack against blackbox next-item recommendation system. The poisoning attack problem is formulated as a multi-step decision problem and is solved via deep reinforcement learning method. In practice, this task could be further complicated by the huge scale of recommendation dataset, the costly training time, and the access restrictions of real recommendation systems.  The proposed framework leverages the influence approximation technique and the recommender simulator. 
Experimental results indicate that the proposed framework consistently outperforms all the baselines in terms of promoting the target items to the target users.
We also study the impact of different factors on the poisoning results. 
In the future, we will investigate the defense strategies against the vulnerability discussed in this paper. 
%\begin{acks}
%To Robert, for the bagels and explaining CMYK and color spaces.
%\end{acks}

\begin{acks}
The work was supported in part by the National Science Foundation under Grant NSF CNS-1742847. The views and conclusions contained in this paper are those of the authors and should not be interpreted as representing any funding agency.
\end{acks}


\clearpage
\balance
\bibliographystyle{ACM-Reference-Format}
\bibliography{acmbib}

% 	\appendix
% 	\subfile{appendix}
%

\end{document}